# USE OF SIMILARITY CRITERIA FOR EVALUATING THE THERMOELECTRIC FIGURE OF MERIT OF SUPERLATTICES

*P.V.Gorskiy*

*Institute of Thermoelectricity of the NAS and MES Ukraine,*

*1 Nauky str., Chernivtsi, 58029, Ukraine*

gena.grim@gmail.com

*In this paper, general relationships that allow evaluating the figure of merit of both two-dimensional superlattices and three-dimensional crystals with a quadratic and isotropic energy spectrum of free charge carriers are derived. It is shown that with the same values of the so-called "dimensionless temperature" and scattering coefficient, the figure of merit of superlattices is always lower than that of three-dimensional crystals. The figure of merit of superlattices can become higher that that of three-dimensional crystals if the "dimensionless temperature" of free charge carrier gas and (or) scattering coefficient in them is considerably higher than in three-dimensional crystals.*

## Introduction

The hopes for improvement of the thermoelectric figure of merit of two-dimensional superlattices as compared to that of three-dimensional crystals [1] are mainly related to the fact that with a quadratic energy spectrum of free charge carriers in both of them there will be always found some finite energy range starting with zero, wherein charge carrier density of state in two-dimensional superlattice is higher than in three-dimensional crystal. However, such hopes are ungrounded, because with a quadratic law of charge carrier dispersion the density of states of charge carriers in two-dimensional superlattice is constant, but finite, whereas in three-dimensional crystal it is proportional to a square root of energy. Hence, in three-dimensional crystal, carriers will be readily found with the energy whereby the density of states in three-dimensional crystal will become higher than that in two-dimensional superlattice. Then the whole problem will be on what side and at what distance from the level of chemical potential of charge carrier gas this energy is, because exactly this defines the ratio between the number of "harmful" and "useful" carriers, increasing and reducing, respectively, the thermoelectric figure of merit. Of vital importance is also scattering coefficient or, in other words, power exponent in the law of energy dependence of the relaxation time of charge carrier quasi-pulse (if we use power model). Therefore, the purpose of the present paper is account of all aforementioned factors with a view of precise comparison of the figure of merit of two-dimensional superlattices with that of three-dimensional crystals.

## Calculation of the figure of merit of three-dimensional crystals and two-dimensional superlattices through similarity criteria

By definition, the value of $ZT$ which characterizes the thermoelectric figure of merit and through which the generator efficiency or refrigerator coefficient of performance are expressed, is a dimensionless value. On the assumption that this value is the integral characteristic of free charge carrier subsystem in material, the use of known relations [2] for three-dimensional crystal yields the following expression for it:

$$ZT(r,\eta) = \left[\frac{(2r+5)F_{r+1.5}(\eta)}{(2r+3)F_{r+0.5}(\eta)} - \eta\right]^2 \left\{\frac{(r+3.5)F_{r+2.5}(\eta)}{(r+1.5)F_{r+0.5}(\eta)} - \frac{(r+2.5)^2 F_{r+1.5}^2(\eta)}{(r+1.5)^2 F_{r+0.5}^2(\eta)}\right\}^{-1}, \quad (1)$$

In this formula, $\eta = \zeta/kT$, $\zeta$ – chemical potential of charge carrier subsystem, $k$ – the Boltzmann constant, $T$ – absolute temperature, $r$ – scattering coefficient, $F_n(x)$ – the Fermi integral of index $n$ with argument $x$. Taking into account the difference between the energy dependences of density of states in three-dimensional crystal and two-dimensional superlattice, its thermoelectric figure of merit can be also determined by formula (1) with replacement of $r$ by $r - 0.5$. But the value $\eta$ for three-dimensional crystal and two-dimensional superlattice is determined differently. Namely, if we introduce the dimensionless temperature $t = kT/\zeta_0$, where $\zeta_0$ – the Fermi energy of electron gas with the absolute zero, the relation between parameters $t$ and $\eta$ is given by the following equations:

$$\frac{2}{3}t^{-1.5} = F_{0.5}(\eta), \quad (2)$$

$$\eta = \ln[\exp(t^{-1}) - 1]. \quad (3)$$

Eq. (2) is valid for three-dimensionless crystal and Eq. (3) – for two-dimensional superlattice.

The results of calculation of parameter $\eta$ from equations (2) and (3) and thermoelectric figure of merit from formula (1) are given in Figs.1 and 2. In the construction of plots in Fig.2 it was taken into account that according to general principles of quantum mechanics, scattering coefficient $r$ can vary from -0.5 to 3.5.

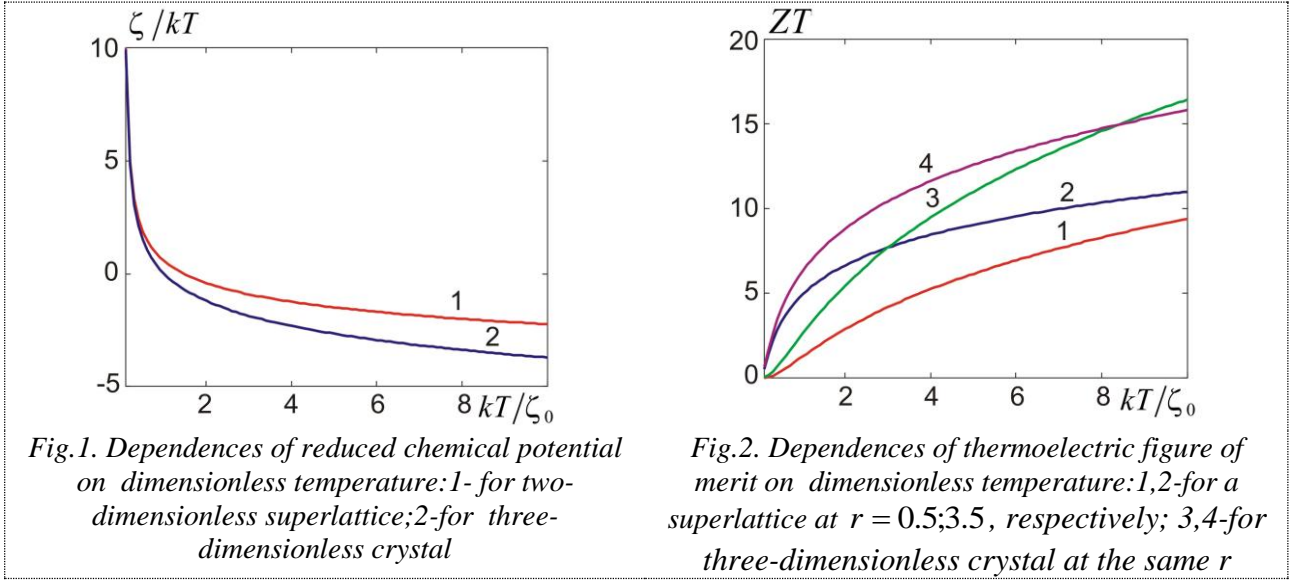

*Fig.1. Dependences of reduced chemical potential on dimensionless temperature: 1- for two-dimensionless superlattice; 2-for three-dimensionless crystal*

*Fig.2. Dependences of thermoelectric figure of merit on dimensionless temperature: 1,2-for a superlattice at $r = 0.5; 3.5$, respectively; 3,4-for three-dimensionless crystal at the same r*

From the figures it is seen that a reduced chemical potential of charge carriers at the same reduced temperature in three-dimensional crystals is lower, and the figure of merit at all $r$ values is higher than in superlattices. Let us illustrate this by some specific numerical examples. Consider a three-dimensional crystal with the bulk charge carrier concentration $n_0 = 3 \cdot 10^{19}$ cm$^{-3}$ and density-of-state effective mass $m^* = m_0$ at $r = -0.5$. For it, at 300K the ratio $t = kT/\zeta_0$ is 19.14, and the thermoelectric figure of merit $ZT = 19$. If, however, we consider a two-dimensional superlattice with the same charge carrier concentration and band parameters and the distance between layers equal to 3nm, for it the ratio $t = kT/\zeta_0$ at 300K will make 1.197, and the thermoelectric figure of merit $ZT = 1.613$. With regard to the contribution of lattice thermal conductivity, this value should be reduced by a factor of about 2.1, which will yield $ZT = 0.77$. This value practically coincides with the experimentally observed, for instance, for thermoelectric material of composition $Bi_{0.5}Sb_{1.5}Te_3$ [2] that has a pronounced layered structure. It is considered that the band structure of this material is rather adequately described by a six-ellipsoid Drabble-Wolfe model [1,3], but the ellipsoids are very oblate in one of directions of layers plane. Therefore, replacement of such surface by an equivalent sphere with density-of-state effective mass, justified in the calculation of the Seebeck coefficient [2], is still very rough in the calculation of the figure of merit. In this case it is more justified to replace a real surface by a cylinder equivalent in charge carrier concentration with its axis normal to layers. This proves the earlier assumption put forward by a number of authors [4-6], including the present author, that conventional thermoelectric materials with a layered structure is a kind of superlattice materials, but they do not possess the set of parameters needed to achieve high values of thermoelectric figure of merit. To achieve such values, it is necessary to have special superlattice thermoelectric materials with low charge carrier concentrations, narrow completely filled allowed and wide forbidden mini bands that must be specially created [6]. From Fig.2 it follows that at $kT/\zeta_0 \leq 2$ it can be done, for instance, by creating special barriers "filtering out" high-energy charge carriers and increasing scattering coefficient.

### Conclusions
1. Thermoelectric figure of merit of materials with a quadratic and isotropic band spectrum of charge carriers, both for three-dimensional crystals and two-dimensional superlattices, increases with a rise in dimensionless temperature and scattering coefficient. In so doing, the figure of merit of superlattices with the same values of dimensionless temperature and charge carrier concentration is less than that of three-dimensional crystals.
2. Traditional thermoelectric materials with a layered structure are also superlattice materials, but they have no combination of parameters needed to reach the high figure of merit.
3. Superlattice thermoelectric materials with the high figure of merit must be developed and manufactured.